\documentclass[journal]{IEEEtran}

\usepackage{graphicx,textcomp}
\usepackage[cmex10]{amsmath,mathtools}
\usepackage{amssymb}
\usepackage{hhline,color,siunitx,xcolor}
\usepackage{blindtext}
\usepackage{esint}
\usepackage{url}

\usepackage[noadjust]{cite}

\newlength\figureheight
\newlength\figurewidth
\begin{document}
\newcommand{\FG}{\color{red}}
\newcommand{\BV}{\color{blue}}
\newcommand{\SE}{\color{green!60!black}}

  \renewcommand{\arraystretch}{1.3}

\title{Current Patterns and Loss Contributions in CORT Cables Carrying AC Current}

\author{B. Vanderheyden, J. Dular, L. Denis, S. Elschner, C. Geuzaine,
  M. Wozniak, F. Grilli \thanks{Document written on \today.}
  \thanks{B.~Vanderheyden, J.~Dular, L.~Denis, and C.~Geuzaine are
    with the University of Liège, Liège, Belgium. S.~Elschner is with
    the University of Applied Science Mannheim, Mannheim,
    Germany. M.~Wozniak is with CERN, Geneva,
    Switzerland. F.~Grilli is with the Karlsruhe Institute of Technology,
    Karslruhe, Germany.}  \thanks{Corresponding author's email:
    B.Vanderheyden@uliege.be} }

\maketitle

\begin{abstract}

  Conductor-on-round-tube (CORT) cables are a potential solution for
  carrying AC power in a small cross-section. Due to the geometry of
  the cable and the helical arrangement of the coated conductors (CC),
  the current follows a non-trivial pattern inside each CC. For
  instance, for the case of a single-layer cable, the current flow is
  mostly axial along the outer face of the CCs and mostly azimuthal
  along their inner face. Such a current distribution, known as the
  \emph{Garber current pattern}, affects the transport AC losses. In
  numerical models, commonly adopted simplifications are either based
  on straight conductors or infinitely thin CCs. Such approaches
  neglect the Garber current pattern and thus misrepresent both the
  detailed current flow within the CC and the resulting 3D
  distribution of the fields.  In this work, the detailed 3D current
  distribution in the CCs is investigated in a one-layer CORT cable,
  as a function of the cable geometrical parameters such as the
  conductor thickness, the pitch angle, and the gap between adjacent
  CCs. In particular, the impact of the Garber current pattern is
  studied on the two largest contributions to the AC losses, namely
  the surface losses (associated with the penetration of the component
  of the magnetic field parallel to the wide faces of the
  superconducting layer) and the edge losses (associated with the
  penetration of the perpendicular component of the magnetic field
  occurring in the vicinity of the gaps between the CCs). The
  detailed distribution of the currents in the CCs is examined and
  its relationship with the different AC loss mechanisms is
  established. This study is carried out by means of an effective 2D
  model that uses a system of coordinates conforming with the helical
  structure of the cable.  The model uses a power-law electrical
  resistivity and for a power index $n = 25$ reproduces the AC losses
  predictions from critical state models to within 40\% for vanishing
  gaps and less than 8\% for straight cables. Calculated AC losses
  converge to the critical state predictions as the value of $n$ is
  increased.

\end{abstract}


\IEEEpeerreviewmaketitle

\section{Introduction}

\IEEEPARstart{W}{ith} their compact arrangement of the tapes and their
mechanical flexibility, conductor on-round-tube (CORT) cables are a
potentially interesting solution for transporting AC power using
high-temperature superconductors (HTS) in the form of coated
conductors (CC), such as in the CORC\textregistered{}
cables~\cite{van_der_Laan:SUST2019}. Possible applications include
power transmission in electric aircraft~\cite{Otten:TAS24} and
underground cables, similarly to those realized with Bi-2223 HTS
tapes, see for example Refs.~\cite{Stemmle:PES14} and
\cite{Willen:TAS25}. CORT cables can also carry high current under
strong magnetic fields and be used in accelerator and fusion
magnets~\cite{Weiss_2020}.  In the design phase, a key aspect is the
evaluation of AC losses. Given the helical structure of the
geometry~\cite{Elschner:TAS15}, the question arises about the
availability and complexity of analytical and numerical models able to
correctly evaluate these losses.

Two analytical models from the literature, the
monoblock~\cite{Vellego:SST95} and the Mawatari~\cite{Mawatari:PRB09}
models, provide first-order approximations in specific conditions. In
the monoblock model, the cable is made of a single piece of
superconductor and the losses are evaluated from the penetration of
the parallel component of the magnetic flux density from the outer
surface. No gap is accounted for. In the Mawatari model, the CCs are
arranged over a cylindrical former and the AC losses are evaluated in
the limit of an infinitesimal thickness. While the gap between each CC
is taken into account, the twist of the CCs is
neglected. Moreover, by considering an infinitesimal thickness, only
the currents induced by the variations of the normal component of the
magnetic flux density contribute to the AC losses. Although both
approaches are useful for first-order estimations, none of them can
capture the complex current patterns arising in CORT cables: as
emphasized in Garber {\it et al.}~\cite{Garber76}, the helical
arrangement of the CCs leads to a non-trivial distribution of the
magnetic flux density, with inner and outer fields in crossed
directions. In the case of a single-layer cable, this situation leads
to a helical current path along the CCs~\cite{Clem:SST10}, i.e. a
zig-zag pattern, with the current flowing along the azimuthal
direction on the CC side facing the inside of the cable and flowing
along the cable axis on the CC side facing the outside of the
cable. Such a current flow, which we call here the Garber current
pattern, needs to be accounted for when evaluating AC
losses~\cite{Clem:SST10}. In Ref.~\cite{Elschner:TAS24}, a 3D
finite element model was used to confirm the existence of the Garber
current pattern and attempt the evaluation of the corresponding AC
losses~\cite{Elschner:TAS24}. The calculations were limited to
artificially inflated thicknesses (to $\SI{400}{\micro\metre}$) due to
computational constraints.  There is therefore a need for a model that
can take into account the combined influence of the gap size, the CC
thickness and the CC pitch angle on AC losses with realistic
parameters, while reproducing the Garber current pattern.

In this work, we use an efficient FEM model to simulate a CORT cable
made of a single layer of three CCs with a realistic
$\SI{1}{\micro\meter}$ thickness.  The model is based on a change of
coordinates that exploits the natural symmetry of the problem and
allows simulating the cable in a 2D geometry~\cite{Dular:TAS24}, thus
considerably reducing the size of the problem and the computation
time. It can model realistic geometries with a large aspect ratio
  of the CC widths to their thickness while properly taking into
  account the helical structure of the single-layer cable.  Throughout
the article, this model is referred to as the {\it helicoidal model}.

The paper is organized
as follows: in section~\ref{s:AC-losses}, the Garber current pattern
arising in helical cables is recalled, together with the different
contributions to AC losses. The helicoidal model is
described in section~{\ref{s:helicoidal-model}}, while the resulting current
distribution and the AC losses are discussed as a function of the gap
size and the twist pitch angle in
section~\ref{s:results}. Section~\ref{s:conclusion} summarizes the
results.

\section{\label{s:AC-losses}AC Loss Contributions in CORT Cables}
\begin{figure}
     \centering
     \includegraphics[width=6 cm]{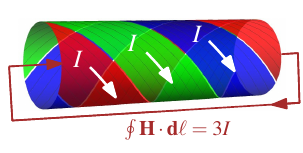}\\[-0.5em]
     (a)\\[1em]
     \includegraphics[width=6 cm]{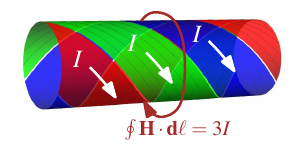}\\[-0.5em]
     (b)\\[1em]
     \includegraphics[width=6 cm]{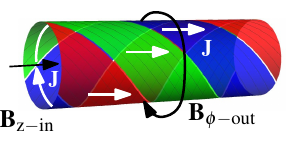}\\[-0.5em]
     (c)
     \caption{Application of Ampere's law to a CORT cable with three
       CCs with a very small gap. (a) The magnetic field produced
       inside resembles that of a long solenoid. (b) The magnetic
       field produced outside resembles that of a long wire. (c) The
       resulting magnetic flux density is axial inside and azimuthal
       outside. Thus, the CCs carry an azimuthal current density on
       their inner face and an axial current density on their outer
       face.}
         \label{fig:Benoit}
\end{figure}

In CORT cables, the CCs are helically wound around a cylindrical
former, as shown in Fig.~\ref{fig:Benoit}. Consider first the limit
where the gap between the CCs is negligible. If one applies Ampere's
law to the two different paths shown in the figure, one can easily see
that, in the inner part of the cable, the magnetic field is axial,
i.e. the cable behaves similarly to a long solenoid, whereas in the
outer part of the cable, the magnetic field is azimuthal, i.e. the
cable behaves similarly to a long straight wire. Now, if one
considers a small finite gap between the CCs, the magnetic field
distribution is nearly the same, except near the CC edges where
fringe fields leak out through the gaps.

The detailed current flow can be derived from the magnetic field
distribution: the current density is azimuthal on the inner surface of
the CCs while it is axial on the outer one, see~Fig.~\ref{fig:Benoit}
(c). By continuity, when reaching the CC edges, the current density
must flow across the CC thickness in order to change from an azimuthal
to an axial direction. The repeated changes of direction results in a
specific zig-zag current trajectory, i.e., the \emph{Garber
    current pattern}.

Based on the description above, one can distinguish three
contributions to the AC losses, as follows.

The first contribution is predominantly in regions away from the gaps.
Although being differently oriented inside and outside the CORT cable, the
magnetic field is parallel to the wide faces of the CCs. Its
penetration inside the superconductor contributes to the so-called
{\it surface losses}. In the framework of the critical state model
(CSM), the surface losses per cycle (in \SI{}{\joule\per\meter\squared}) are
given by equation (3.11) of~\cite{Campbell:AP72}
\begin{equation}\label{eq:surface}
	Q_{\rm surface}=\frac{2 B^3}{3 \mu_0^2 J_{\rm c}}
\end{equation}
where $B$ is the magnetic flux density parallel to the surface (in the
CORT case, created by the transport current), $\mu_0$ is the vacuum
permeability, and $J_{\rm c}$ is the superconductor critical current
density, here assumed to be independent of $B$.  This well-known formula
describes the losses occurring at both of the wide faces of the CCs, so that the
total surface loss is the sum of inner and outer contributions,
see~\cite{Elschner:TAS24}.  The result is valid only below the
flux-cutting threshold, i.e. when the magnetic flux fronts penetrating
inside the CCs have not joined together.  Above this limit, the
penetrating magnetic flux fronts overlap and the resulting interaction
between the field components leads to complex current profiles across
the thickness of the layer. To our knowledge, no analytical description
is available in this case.

The second loss contribution arises in regions close to the gap between
the CCs. These losses are called {\it edge losses} and are caused by
the penetration of the magnetic field perpendicular to the wide face
of the CCs.  In~\cite{Mawatari:PRB09}, Mawatari provided an expression
for the AC losses (per CC, in \SI{}{\joule\per\meter}) of curved
CCs arranged conformally to a cylinder as
\begin{equation}\label{eq:Mawatari}
Q_{\rm edge}=\frac{\mu_0 I_{\rm c}^2}{\pi}F^2 \int_0^1 (1-2s) \ln \left [ 1-\frac{\tan^2(F\,s\,\theta_n)}{\tan^2 \theta_n}\right ] {\rm d} s
\end{equation}
where~$F=I_{0}/I_{\rm c}$ is the ratio between the amplitude of the
transport current and the critical current
($I_{\rm c} = J_{\rm c} w d$, with $w$ the CC width and $d$ its
thickness), while $\theta_n=N_t w/4R$ with $N_t$ the number of CCs and
$R$ the cylinder radius.  In the Mawatari model, each CC is
approximated as an infinitely thin superconducting layer. This means
that, in the superconductor, only the magnetic field component
perpendicular to the wide face of the CC is taken into
account. Additional information on the physics behind this assumption
can be found in Brandt's papers, for example in~\cite{Brandt:PRB93}.

Finally, the third loss contribution is given by the currents connecting
the inner azimuthal and outer axial current flows along the CCs,
which cause a magnetic field parallel to the CC edges.  In the rest
of this paper, these losses will be referred to as \emph{diving current
  losses}. To the best of our knowledge, there is no analytical
expression for this loss contribution.

In~\cite{Elschner:TAS24}, a 3D finite element method (FEM) was used to
model to confirm the presence of the Garber current pattern in a CORT
cable composed of three HTS coated conductors. In order to keep the
size of the problem and the computation time reasonable, the thickness
of the superconductor was increased from a realistic value
of~\SI{1}{\micro\meter} to~\SI{400}{\micro\meter}. The critical
current density $J_{\rm c}$ was reduced by the same factor 400, in
order to keep $I_{\rm c}$ fixed.

This artificial expansion of the thickness of the superconductor
introduces a problem for the calculation of the AC losses. In the FEM
model, the power dissipation is calculated by integrating
$\bf J \cdot \bf E$, where $\bf E$ is the electric field, over the
superconducting domain. In the case of a superconductor with an
artificially expanded thickness, the question arises whether the
results obtained from the integration should be rescaled by a factor
depending on the thickness. If the dissipation were purely due to
surface losses, the resulting losses should be divided by the
expansion factor: as shown in ~\eqref{eq:surface}, the surface losses
are inversely proportional to $J_{\rm c}$ (which is reduced in the
artificially expanded superconductor). If, on the other hand, the
dissipation were purely due to the edge losses, the results of the
integration should not be rescaled as the thickness is not involved
  in Mawatari's equation~\eqref{eq:Mawatari}. Only the critical
current, the CC width, and the gap matter. In the case of an
artificially expanded thickness, all these parameters remain the same
as in the realistic case and the losses do not change. Last, if the
dissipation were purely due to the diving current losses, in the
absence of any analytical evaluation it would be impossible to know
whether a scaling law applies.

The dominance of one loss contribution over the other depends on
parameters such as the thickness, the gap between CCs, the twist
angle, and the amplitude of the transport current. The Garber current
pattern further complicates matters, because it introduces radial
currents (and loss components), which are neither considered
in~\eqref{eq:surface} nor in~\eqref{eq:Mawatari}.  Hence, in order to
make an accurate estimation of the losses, one needs a numerical model
that is able to simulate the helical structure of the CORT cable with 
the actual physical parameters.

\section{\label{s:helicoidal-model}Helicoidal Model} 

\begin{figure}
     \centering
     \includegraphics[width=0.95\linewidth]{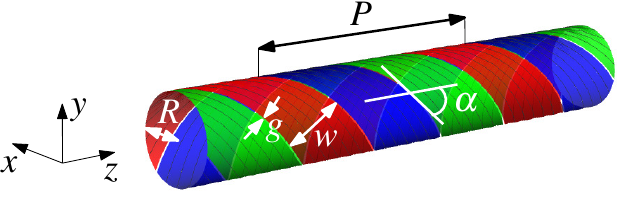}\\[-0.5em]
     (a)\\[1em]
     \includegraphics[width=0.95\linewidth]{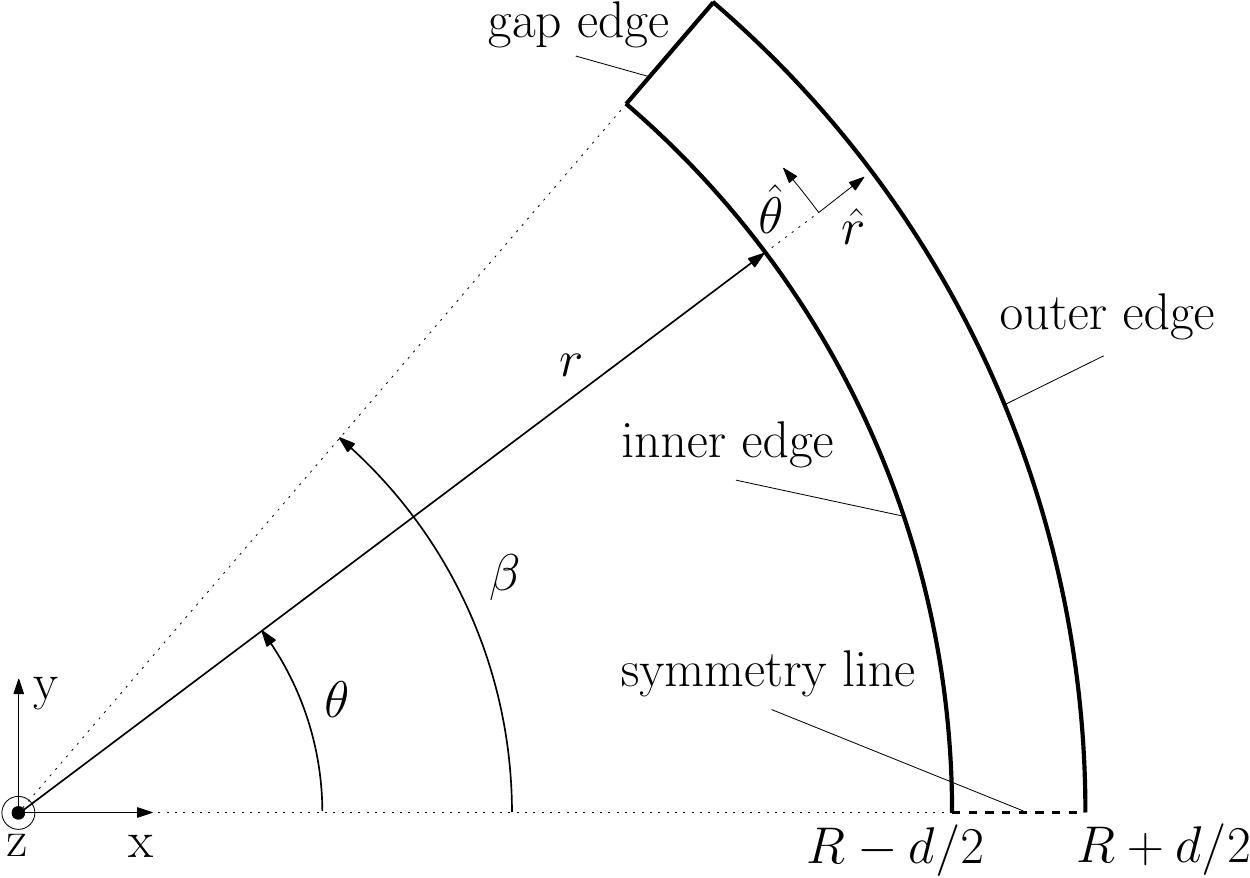}\\[-0.5em]
     (b)\\[1em]
     \caption{Geometry of the single-layer CORT cable and parameter
       definition. (a) 3D-view, where $R$ is the mean radius, $g$ the
       width of the gap between CCs, $w$ the CC width, $\alpha$
       the pitch angle, and $P$ the pitch length. (b) Cross section at
       a fixed value of $z$. Only one half of a CC is shown and the
       CC is artificially enlarged for clarity. Here, $d$ is the
       CC thickness, $r$ and $\theta$ are cylindrical coordinates
       and $\beta$ is half the angular size of the CC as seen from
       the $z$-axis in the $x$-$y$ plane.}
         \label{fig:3D+cross-section-definitions}
\end{figure} 

In this work, we model the current density distribution, field
distribution, and AC losses in a CORT cable transporting a sinusoidal
current. The cable is made of a single layer of three CCs of width
$w$ and thickness $d$, wound around a former tube of radius $R - d/2$
with a pitch angle $\alpha$ and a separation gap $g$, as shown in
Fig.~\ref{fig:3D+cross-section-definitions} (a). The pitch
(periodicity) length $P$ is given as~\cite{Elschner:TAS24}
\begin{equation}
  \label{eq:parameter-relationships}
  P =  \frac{2 \pi R}{\tan\alpha},
\end{equation}
with the mean radius for three CCs ($N_t = 3$) given as 
\begin{equation}
  \label{eq:mean-radius}
  R = \frac{3(w + g)}{2 \pi \cos\alpha}.
\end{equation}

The magnetic response of the cable is calculated using the helicoidal
model of Dular {\it et al.}~\cite{Dular:TAS24}, which is based on an
$H$-$\phi$ formulation and exploits the helical symmetry of the
cable. For the case of CCs transporting currents with no transverse
applied field, the field variables are invariant along the tube axis
when expressed in a coordinate system conforming to the helical
geometry of the cable. As a result, the problem can be solved directly
in the 2D-geometry of the cable cross section. However, note that the
$H$-field has three components, so that the model is effectively
describing a 3D geometry even if its support is a 2D domain. The
superconducting layers are described with the power-law resistivity
\begin{equation}\label{eq:power-law}
	\rho(J)=\frac{E_{\rm c}}{J_{\rm c}}\left( \frac{J}{J_{\rm c}}\right)^{n-1},
\end{equation}
where $J$ is the current density modulus,
$E_{\rm c}=\SI{1e-4}{\volt\per\meter}$, $J_{\rm c}$ is the critical
current density and $n$ is the power-law index. In this work, we
consider $J_{\rm c}$ to be independent of the magnetic field magnitude
and orientation. As the origin of the Garber effect is geometrical,
this assumption is not restrictive with respect to the physical effect
under study.

In the remainder of this section, we focus on the definition of the
geometrical and operational parameters, while the simulation
parameters are further detailed in Appendix~\ref{A:helicoidal-model}.
Figure~\ref{fig:3D+cross-section-definitions} (b) shows the
cross-section of a single CC in the $x$-$y$ plane (only one-half of
the CC is depicted). For a 3-CC configuration, The angular size
$\beta$ is given as
\begin{equation}
  \label{eq:angular-size}
  \beta = \frac{w}{2 R \cos\alpha} = \frac{\pi}{3}\,\frac{w}{w+g}.
\end{equation}
Each of the three CCs is \SI{4}{\milli\meter} wide,
\SI{1}{\micro\meter} thick, and carries a sinusoidal current
$I(t) = I_{0} \,\sin(2 \pi f t)$ of frequency $f = 50~$Hz (the CC
currents are in phase). The critical current density is taken as
$J_{\rm c}=\SI{4e10}{\ampere\per\meter^2}$ and the single-CC
critical current is $I_{\rm c} = J_{\rm c} \,w \,d = \SI{160}{\ampere}$.
In order to remain under the flux-cutting regime, the current
amplitude is chosen to be $I_0 = \SI{90}{\ampere} < I_{\rm th}$, where
\begin{equation}
  \label{eq:flux-cutting-threshold}
  I_{\rm th} = \frac{I_{\rm c}}{\cos\alpha+\sin\alpha}
\end{equation}
is the flux-cutting current threshold~\cite{Elschner:TAS24}. As will be
detailed below, this condition leads to separate magnetic flux fronts
penetrating from the inner and outer faces of the CCs. In contrast,
the regime with $I_{0} \ge I_{\rm th}$ leads to overlapping magnetic flux
fronts inside the CC; this flux-cutting regime will be addressed in
a separate work.

\section{\label{s:results}Results}
The current distribution and the AC loss mechanisms are analyzed for a
single-layer CORT cable while varying its geometrical parameters.
When varying the gap $g$, attention should be paid to its influence on
the other parameters. According to~\eqref{eq:mean-radius}, varying $g$
while keeping $\alpha$ fixed also affects the radius $R$. From
Ampere's law, this affects the $B$-field component parallel to the CC
outer faces and hence the surface AC losses. Thus, variations of $g$
cannot be considered as a means to probe gap losses only.  One may be
tempted to counterbalance the change of the magnetic flux density by
adapting the amplitude $I_{0}$ of the current carried by the CCs,
but this choice then affects the critical ratio $F=I_{0}/I_{\rm c}$,
which also impacts the gap losses, see ~\eqref{eq:Mawatari}. This
parameter interdependence complicates a proper understanding of the
loss mechanisms, as was already emphasized in~\cite{Siahrang:TAS10}.
Here, in order to focus on the effect of the helical structure of the
CORT cable, we opted for keeping a fixed mean radius $R$ while
varying both the gap $g$ and pitch angle $\alpha$. The mean radius was
chosen for a reference geometry with a CC thickness
$d = \SI{1}{\micro\meter}$, a CC width $w = \SI{4}{mm}$, a gap
$g = \SI{0.1}{mm}$ and a pitch angle $\alpha =
\pi/4$. Following~\eqref{eq:mean-radius}, this yields
$R = \SI{2.768}{mm}$ and, for a given gap $g$, a pitch angle
\begin{equation}
  \label{eq:alpha-vs-gap}
  \alpha(g) = \cos^{-1}\left(\frac{3(w+g)}{2 \pi R}\right).
\end{equation}

\subsection{\label{ss:current-density-distributions}Current density
  distributions}

\begin{figure}
  \centering
  \includegraphics[width=0.4\linewidth]{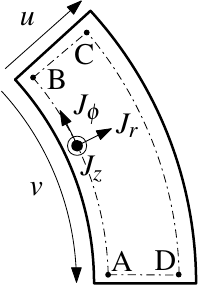}\quad%
  \includegraphics[width=0.5\linewidth]{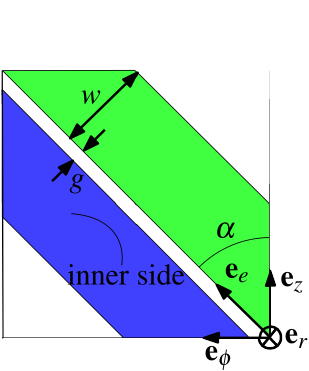}
  \caption{Various definitions for representing the current
    distribution in the CC. Left: half-CC viewed along the cable axis from
    above. Paths along which the current density is represented: AB,
    along the inner edge, BC, along the gap edge, CD, along the outer
    edge, and DA, along the conductor symmetry line (see
    Fig.~\ref{fig:3D+cross-section-definitions} (b)). Here, $J_z$,
    $J_\phi$, and $J_r$ are respectively the axial, azimuthal, and
    radial components of the current density. Relative variables
    $u \in [0,1]$ and $v \in [0,1]$ are used, with $(0,0)$
    representing the gap vertex on the inner CC edge and $(1,1)$ the
    intersection of the conductor symmetry line and the outer CC
    edge. Right: unrolled plot of two CCs,
    see Fig.~\ref{fig:3D+cross-section-definitions} (a). Definition of unit
    vectors: ${\bf e}_r$, ${\bf e}_\phi$, and ${\bf e}_z$ respectively
    point along the radial, azimuthal, and axial direction, while
    ${\bf e}_e$ is perpendicular to the radial direction and points
    along the CC edges.}
  \label{fig:cort-paths-4-plots}
\end{figure}

We now turn to analyzing the distribution of current densities as a
function of the gap size, for a fixed mean radius $R$. The current
densities are evaluated at a quarter period $t=\SI{5}{\milli\second}$,
i.e. when the current $I(t)$ reaches its maximum for the first time.
In the critical state limit, this situation would correspond to the
maximum penetration of the magnetic flux in the
CCs. Figure~\ref{fig:cort-paths-4-plots} (left) illustrates the
different paths along which the current density is probed and defines
the relative variables $u$ and $v$,
  \begin{equation}
    \label{eq:uv}
  u = \frac{r - (R - d/2)}{d},\quad
  v = 1 - \frac{\theta}{\beta},
\end{equation}
varying continuously from $0$ to $1$ as the position is changed,
respectively, along a radial line from the inner to the outer edge of
the CC and along the azimuth from the gap edge to the CC symmetry
line.  The origin $(u,v)=(0,0)$ is located at the intersection of
  the gap edge and the inner CC edge, while the point $(1,1)$ is
  located at the intersection of the symmetry line and the CC outer
  edge.  Simulations were carried out on one half of a CC with a fine
structured mesh containing 20 elements across the thickness and 420
elements along the azimuth, as described in
Appendix~\ref{A:helicoidal-model}. Paths AB and CD respectively
correspond to rows $\#1$ and $\#20$ along the thickness, whereas paths
BC and DA correspond to rows $\#1$ and $\#420$ along the
azimuth. To understand better the 3D distribution of current
densities, Fig.~\ref{fig:cort-paths-4-plots} (right) shows the
cylindrical unit vectors ${\bf e}_r$, ${\bf e}_\phi$, and ${\bf e}_z$,
together with the edge unit vector ${\bf e}_e$, that is parallel to
the CC edge and is perpendicular to ${\bf e}_r$.

\begin{figure*}
  \centering
  \includegraphics[width=\linewidth]{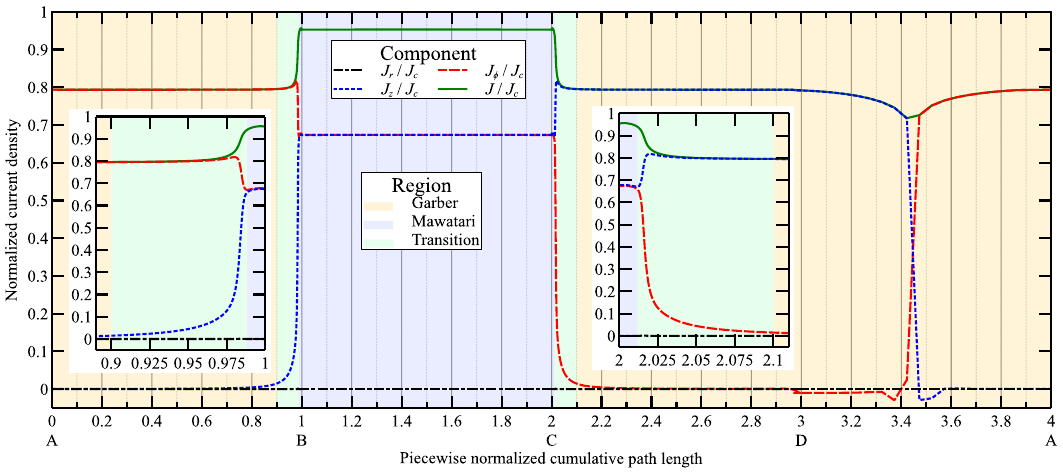}
  \caption{Current density in one half of a CC for the reference
    geometry with a gap of $\SI{100}{\micro\meter}$ and $n =
    25$. The components of the current density are represented along
    the loop ABCDA as a function of the cumulative path index, which
    is normalized in such a way that it increases by one unit along
    each leg of the loop. The Garber region is colored in yellow and
    exhibits a current density that is axial near the outer face of
    the CC and azimuthal near the inner face of the CC. The Mawatari
    region is colored in lavender and exhibits a current density that
    flows along the edge unit vector ${\bf e}_e$, in this case
     with $J_\phi=J_z$ ($\alpha = \pi/4$). The transition regions are colored in
    green and continuously connect the Garber and Mawatari regions. The
    insets are close-ups of the distribution of the current density
    near B and C and show the presence of the Mawatari region over a small part of the paths AB and CD.}
  \label{fig:continuous_path_n_25}
\end{figure*}

Figure~\ref{fig:continuous_path_n_25} shows the current density
components for the reference geometry, with a gap of
$\SI{100}{\micro\meter}$, corresponding to a pitch angle
$\alpha = \pi/4$, and $n = 25$. The current density components are
shown along a continuous loop ABCDA as a function of the piecewise
cumulative path length $P_L$, that is normalized so that it is
incremented by one unit after the completion of each leg of the
loop. Three different regions can be distinguished. First, starting
along AB, the current density is azimuthal. It remains azimuthal over
most of the path AB and is axial over most of the path CD, thus
following the Garber current pattern. This \emph{Garber region},
represented in yellow, is present over a large azimuthal extent of the
CC. A second region can be distinguished near the gap at $v = 0$
(i.e. near B and C), where the components $J_\phi$ and $J_z$ are
equal, while the radial current is negligible. This is observed along
BC as well as along a short length of both AB and CD as shown in the
insets.  Here, the equality of $J_\phi$ and $J_z$ is a result of the
cable helical geometry. With a pitch angle $\alpha = \pi/4$, the
current density is in fact flowing along the edge direction
${\bf e}_e = ({\bf e}_z+{\bf e}_\phi)/\sqrt{2}$, so that one recovers
in this region the current pattern predicted by the Mawatari
model. This \emph{Mawatari region} is represented in lavender.  The
last identified region is a \emph{transition region}, represented in
green, which continuously connects the Mawatari and Garber
regions. Last, as one closes the ABCDA loop along path DA, one finds
again the Garber current pattern: the current density is purely axial
near D and switches to a purely azimuthal current density near
A\footnote{It can be observed that the plot showing the current
  density along DA is slightly shifted to the left. This negative
  shift of the cumulative path length arises from the mesh partition
  of the conductor, which is made of a collection of trapezoids
  distributed along a circular path, with the innermost parallel edges making
  a chord. As a result, the corresponding range of the reduced
  variable $u$ is shifted to negative values, the effect being
  stronger near the symmetry line where the elements are longer.}. The
two current fronts are barely touching at the middle of DA. Note that
the simulations show oscillations in the $J_\phi$ and $J_z$ components
with negative values ahead of the fronts. To the best of our
knowledge, such oscillations are numerical artefacts akin to those
observed in simulations of a magnetic flux diffusing in a
one-dimensional slab~\cite{Dular-TAS20}. Oscillations occur when the
magnetic flux diffuses in an initially virgin region and are reduced
with a refinement of the mesh.

Interestingly, when comparing the Garber and Mawatari regions, it can
be observed that the modulus of the induced current density $J$ is
different in each region, emphasizing the fact that the magnetic flux
penetration mechanisms are distinct and so are the time variations of
the magnetic flux and the induced electric fields.  For the chosen
frequency of the current waveform $I(t)$, the current density $J$ is
subcritical, as discussed further below.  Note that for representing
the current density along the loop, it should be kept in mind that
very different physical lengths are involved: the azimuthal path
length of the CC is $\SI{2.8}{mm}$, while its thickness is
$\SI{1}{\micro\meter}$.

\begin{figure}
  \centering
  \includegraphics[width=\linewidth]{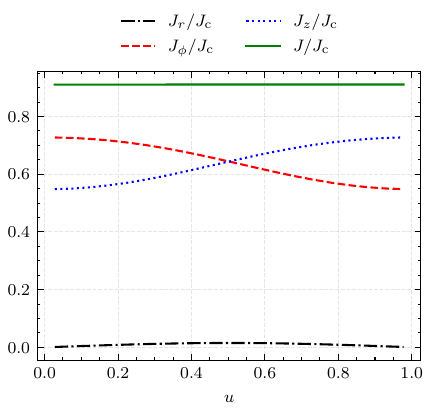}
  \caption{Current density components along a radial path within the transition region ($v = 0.015$). Reference geometry with  $\alpha = \pi/4$, $g = \SI{100}{\micro\meter}$, and $n = 25$.  }
  \label{fig:PMo_n_25}
\end{figure}

\begin{figure}
  \centering
  \includegraphics[width=\linewidth]{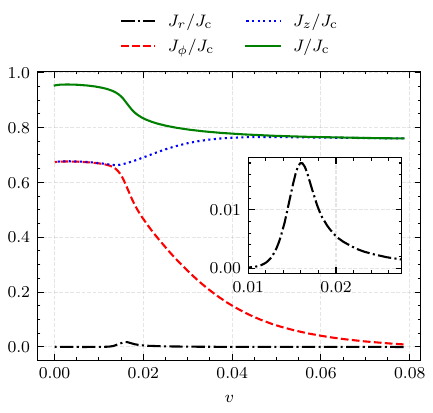}
  \caption{Current density along an azimuthal path near the middle of the CC. Reference geometry $\alpha = \pi/4$, $g = \SI{100}{\micro\meter}$, and $n = 25$. The selected path is along row \#11.}
  \label{fig:Pmed_n_25}
\end{figure}

 Last, a few additional plots complete the picture on the transition
 between the different regions: Fig.~\ref{fig:PMo_n_25} shows the
 current density along a radial path in the transition region
 ($v = 0.015$). The $J_\phi$ and $J_z$ distribution are seen to evolve
 to a Garber pattern, as for values of $u$ near $0$,  $J_z$ is
 depressed with respect to $J_\phi$ (as compared to the Mawatari
 region where $J_z = J_\phi$), while the opposite is observed
 for values of $u$ near $1$. In parallel, the radial component $J_r$
  slightly raises with a maximum at $u \simeq 0.5$
 or $r \simeq R$. Figure~\ref{fig:Pmed_n_25} shows the current density
 along an azimuthal path near the middle of the CC (row \#11,
 $u = 0.57$). The radial current density is seen to be negligible
 almost everywhere except at the interface between the Garber and the
 transition region where it peaks at
 $J_{r-{\rm max}} = 1.8\times 10^{-2}~J_{\rm c}$.

\begin{figure*}[h]
  \centering
  \includegraphics[width=\linewidth]{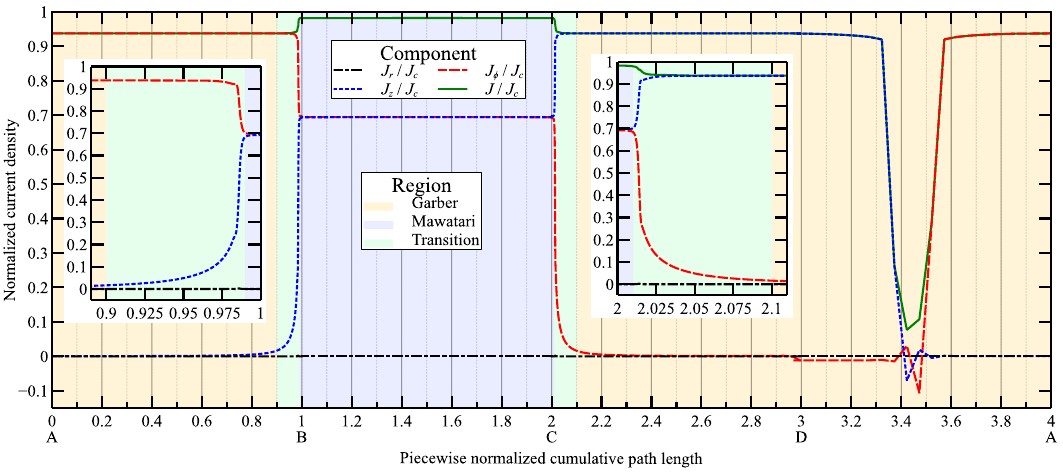}
  \caption{Current density in one half of a CC for the reference
    geometry with a gap of $\SI{100}{\micro\meter}$ and $n = 101$. The
    layout of the plot uses the same conventions in
    Fig.~\ref{fig:continuous_path_n_25}. Here, one can again
    distinguish the three regions --- Garber, Mawatari, and transition
    regions ---. In contrast to the case $n = 25$, the current density
    $J$ is larger and gets closer to $J_{\rm c}$ in the Mawatari
    region than in the Garber one. Along the path DA, the flux fronts
    are sharper than those in Fig.~\ref{fig:continuous_path_n_25}.}
  \label{fig:continuous_path_n_101}
\end{figure*}

\begin{figure}
  \centering
  \includegraphics[width=\linewidth]{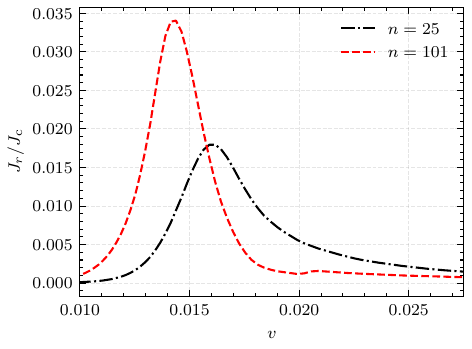}
  \caption{Comparison of the radial component $J_r$ along an azimuthal
    path near the middle of the CC for the reference geometry with a
    gap of \SI{100}{\micro\meter}, for $n = 25$ and $n = 101$. The
    selected path is along row \#11.}
  \label{fig:jr-n-25-vs-101}
\end{figure}

The previous results exhibited subcritical current densities $J$ over
most of the cross section of the CC. Raising the $n$-value to $n = 101$ yields current
distributions with a very similar structure (with a Mawatari, Garber,
and transition regions), but higher levels of induced current density,
as shown in Fig.~\ref{fig:continuous_path_n_101}. Now
$\sim 0.98\times J_{\rm c}$ in the Mawatari region and
$\sim 0.94\times J_{\rm c}$ in the Garber region. (Compare with
Fig.~\ref{fig:continuous_path_n_25} for $n = 25$.)  In particular, it
can be observed that the critical state limit is approached more
slowly as $n$ increases in the Garber region than in the Mawatari
region, this has implications on the sensitivity of the AC losses with
the power-law index $n$, as discussed below. The radial component $J_r$ is also
increased, as shown in Fig.~\ref{fig:jr-n-25-vs-101} along an
azimuthal path drawn through about the middle of the CC. In addition, the
current front separation is more marked in the Garber region, as shown
along the path DA in Fig.~\ref{fig:continuous_path_n_101}, to be
compared with Fig.~\ref{fig:continuous_path_n_25} for $n = 25$.

\begin{figure}
\centering
\includegraphics[width=0.666\linewidth]{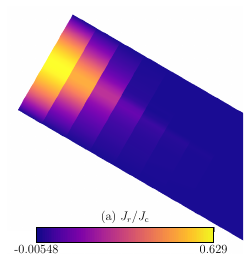}\\
\includegraphics[width=0.666\linewidth]{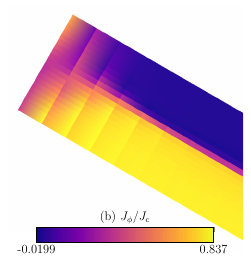}\\
\includegraphics[width=0.666\linewidth]{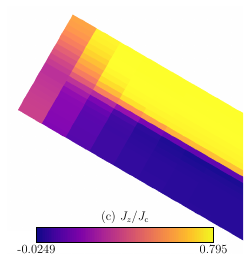}
\caption{Current distribution for a gap of $125~$nm and $n =
  25$. Close-up near the gap edge}
\label{fig:sg_j-near-gap}
\end{figure}

\begin{figure}
\centering
\includegraphics[width=0.6\linewidth]{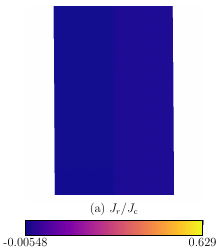}\\
\includegraphics[width=0.6\linewidth]{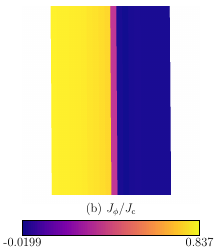}\\
\includegraphics[width=0.6\linewidth]{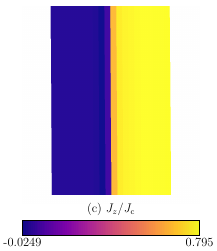}
\caption{Current distribution for a gap of $125~$nm and $n =
  25$. Close-up near the symmetry plane.}
\label{fig:sg_j-near-sym}
\end{figure}

\begin{figure}
  \includegraphics[width=\linewidth]{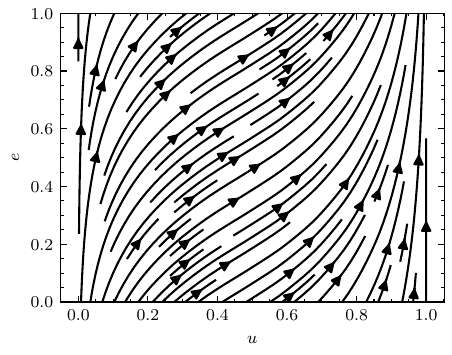}

  \caption{Stream lines of the current density over the edge face of a
    CC for a gap $g=\SI{125}{\nano\meter}$ and $n = 25$. Here, $u$ is
    the relative variable defined in Fig.~\ref{fig:cort-paths-4-plots}
    with $u = 0$ ($u=1$) corresponding to the inner (outer) face of
    the CC, respectively, while $e$ is a relative variable for positions
    along the edge unit vector, ${\bf e}_e$ . The stream lines are shown for a patch of area $1\times 1~\SI{}{\micro\metre}^2$.}
  \label{fig:edge-currents-n-25}
\end{figure}

The pattern of current distribution that we just described can be
observed for most gap values. It is of interest to look at two
limiting cases. First, consider a very small gap
$g = \SI{125}{\nano\meter}$, corresponding to a pitch angle
$\alpha = \SI{0.8095}{\radian}$, or $\SI{46.4}{\degree}$
(see~\eqref{eq:alpha-vs-gap}). The Garber region is found to extend
over most the length of the CC, whereas another region appears near
the gap edge, as illustrated in Figs.~\ref{fig:sg_j-near-gap} and
\ref{fig:sg_j-near-sym}. Very near the gap edge, the current density
${\bf J}$ tends to align with the edge unit vector ${\bf e}_e$. This
is reminiscent of the current density observed in the Mawatari region
for a larger gap $g = \SI{100}{\micro\meter}$ and is related to the
fact that even though the present gap is small, a magnetic flux
density can thread through it. As a result, the magnetic flux density
permeates through the gap inside the CC, with ${\bf B}$ smoothly
changing from an axial direction at the smallest radius to an
azimuthal one at the largest radius.  Further away from the edge, it
can be noticed that the current density has a non-negligible radial
component, over a region extending along the azimuth by
$\sim \SI{1}{\micro\meter}$. This pattern corresponds to diving
currents making a connection between the inner azimuthal flow and the
outer axial flow. To illustrate further the diving currents,
Fig.~\ref{fig:edge-currents-n-25} shows the stream lines of the
current density field over the edge face of the CC, for a patch of
area $\sim 1 \times 1~\SI{}{\micro\meter}^2$. Here, $u$ is the
relative radial coordinate used above, while $e$ is the relative
coordinate along the edge unit vector ${\bf e}_e$. This graph is
constructed on a distribution of the current density that, according
to the helical symmetry of the cable, is independent of $e$ and
coincides for $e = 0$ to the current density along the gap edge of the
CC (i.e., along the line $v = 0$ with $u \in [0,1]$).  It can be
observed that the diving currents are following a non-trivial pattern,
flowing strictly along ${\bf e}_e$ for $u = 0$ and $u = 1$, and along
an inclined direction with a positive radial component for
intermediate values of $u$.

The second limiting case is that of large gaps. As $g$ increases, the
pitch angle decreases and vanishes when the CCs are parallel to the
cable axis. From~\eqref{eq:alpha-vs-gap}, this corresponds to a
maximum gap $g_{\rm max} = 2 \pi R/3 - w = \SI{1.8}{\mm}$. A detailed
analysis of configurations with a gap width $g \lesssim g_{\rm max}$
shows that the induced current densities near the gap follow the
configuration described in the Mawatari model, with ${\bf J}$ aligning
along the edge unit vector ${\bf e}_e$, while a perpendicular magnetic
flux density is penetrating the CCs.  Note however that due to the
finite thickness of the CC, the magnetic flux lines are strongly
curved, with an azimuthal component that changes sign from the inner
to the outer face. In contrast, the currents near the symmetry line
follow the Garber pattern, with an axial component along the outer face
and an azimuthal component along the inner face. With
$g \lesssim g_{\rm max}$, the pitch angle $\alpha$ is small, resulting
in a small net inner Garber current. Moreover, the inner current also
shows an axial component, that is induced by an in-plane magnetic flux
density permeating the cable former region through the large gaps.

\subsection{\label{ss:AC-losses}AC losses}

The complex current distribution observed in the cable for different
geometries has an impact on its transport AC losses.  In the critical
state limit, the Garber current pattern has been shown to yield AC
losses per cable length $L$ (i.e., the losses of the 3 CCs in \SI{}{\joule\per\meter}) given
as~\cite{Elschner:TAS24}
\begin{equation}
  \label{eq:AC-losses-Garber}
  \frac{Q_{\rm Garber}}{L} = \frac{2 \mu_0 I_{0}^3}{J_{\rm c}(w+g)^2}\,\left(\frac{\cos^3\alpha+\sin^3\alpha}{\cos\alpha}\right).
\end{equation}

\begin{figure}
     \centering
           \includegraphics[width=8 cm]{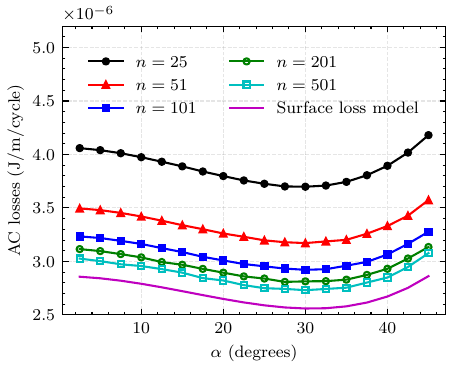}
           \caption{AC losses as a function of the pitch angle, for a
             constant gap of~\SI{0.5}{\micro\meter} and for different
             values of $n$. The surface loss model is
             given in~\eqref{eq:AC-losses-Garber}.}
         \label{AC-losses-vs-pitch}
\end{figure}

Figure~\ref{AC-losses-vs-pitch} shows the AC losses obtained
numerically for the case of a cable with a gap
$g = \SI{0.5}{\micro\meter}$, as a function of the pitch angle
$\alpha$ for different values of $n$. Note that here
the gap is fixed, so that the radius $R$ varies with $\alpha$,
according to~\eqref{eq:mean-radius}. The numerical losses were
evaluated over the second half cycle as 
\begin{equation}
  \label{eq:P(t)}
  \frac{Q_{\rm num}}{L} = 2\,\int_{1/2f}^{1/f} \int {\bf E}\cdot{\bf J}\,dS\,dt,
\end{equation}
with $S$ the full cross section of the cable at fixed $z$.

The different curves follow the
expected angular dependence in~\eqref{eq:AC-losses-Garber}. It can be
noticed that the losses for $n = 25$ are about \SI{40}{\percent}
higher than those predicted in~\eqref{eq:AC-losses-Garber}. Moreover,
the critical state limit predicts identical losses for $\alpha = 0$
and $\alpha = \pi/4$ ($\SI{45}{\degree}$), whereas for $n = 25$, losses for
$\alpha = \pi/4$ are higher than those for $\alpha = 0$. These
differences are reduced as $n$ is increased: both the excess to the
theoretical losses and the difference between $\alpha = 0$ and
$\alpha = \pi/4$ are decreasing as $n$ goes from $25$ to
$501$.

\begin{figure}
     \centering
          \includegraphics[width=\linewidth]{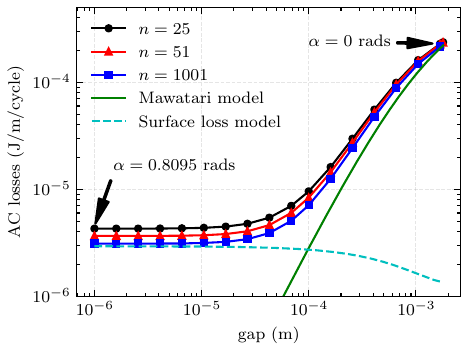}
          \caption{AC losses as a function of the gap for a fixed
            radius and a transport current of \SI{90}{\ampere} per
            CC. Analytical predictions by the surface loss and
            Mawatari models are also shown. Note that the point
            representing the losses for $n = 1001$ and $g=g_{\rm max}$
            is missing due to a convergence issue.}
         \label{ACLossVsGapAgainstModels}
\end{figure}

Figure~\ref{ACLossVsGapAgainstModels} shows the AC losses as a
function of the gap. In this case, the radius is fixed again to
  $R = \SI{2.768}{mm}$ and the pitch angle varies with the gap
  following~\eqref{eq:alpha-vs-gap}, from
  $\alpha = \SI{0.8095}{\radian}$ for $g\to0$ to $\alpha = 0$ for the
  maximum gap $g_{\rm max} = \SI{1.8}{mm}$. The surface loss model
  curve follows~\eqref{eq:AC-losses-Garber}. The surface losses are nearly
  constant for small gaps, but decrease at large gaps due to the
  dependence of $Q_{\rm Garber}/L$ with respect to both $g$ and $\alpha(g)$. The
  Mawatari model curve shows the losses evaluated in a situation assuming
  three infinitely thin CC conformal to a tube of radius $R$, with a
  varying separation gap. The CCs are assumed to be straight.

For very small gaps, the simulated losses are low and follow the
surface losses: there is no dependence on the gap because the edge
losses are practically suppressed. The edge losses start having an
influence for gaps $\geq$~\SI{10}{\micro\meter} and they become
dominant as the gap further increases, converging toward the value
predicted by the Mawatari model. It is worth noting that the
  simulated results exhibit larger deviations at small gaps than at
  large gaps: for $g < \SI{10}{\micro\meter}$, losses for $n = 25$ are
  about $40~\%$ higher than those predicted by the surface loss model,
  while they converge to the critical state limit as $n$ is
  increased. In contrast, at the largest gaps near $g_{\rm max}$, the
  simulated losses show a lower deviation from the Mawatari
  prediction, of less than $8~\%$ for $n = 25$ while the results
  converge to the predicted losses as $n \to \infty$.

\begin{figure}
     \centering
          \includegraphics[width=\linewidth]{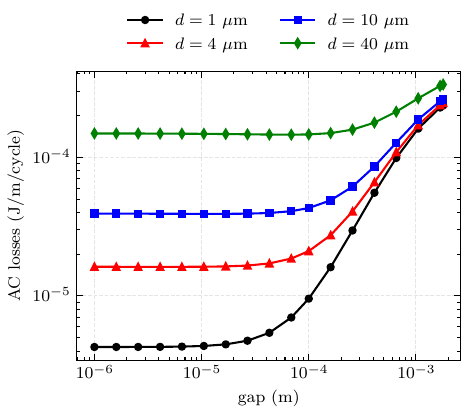}
          \caption{AC losses as a function of the gap for a fixed
            radius and a transport current of \SI{90}{\ampere} per
            CC and different thicknesses of the superconducting
            layer.}
         \label{fig:ACLossVsGapDifferentThickness}
\end{figure}

Figure~\ref{fig:ACLossVsGapDifferentThickness} shows the AC losses as
a function of the gap, calculated for different values of the
thickness of the superconducting layer, with $J_{\rm c}$ scaled so
that the critical current $I_{\rm c}$ is unchanged. For small gap
values, the losses are dominated by the surface loss contribution, and
the values calculated with \eqref{eq:P(t)} scale approximately with
the thickness of the superconductor, according
to~\eqref{eq:surface}. At large gaps (e.g. \SI{1}{\milli\meter}), the
losses are dominated by the edge contribution, and the values
calculated with \eqref{eq:P(t)} are not influenced by the scaling,
according to \eqref{eq:Mawatari}. Actually, this is true as long as
the infinitely thin approximation is applicable. This is already not
the case for a thickness of \SI{40}{\micro\meter}, whose losses are
distinguishably different from those corresponding to a thinner
superconductor.
       
In addition to confirming the expected scaling behavior at small and
large gaps, these results highlight the importance of having a
numerical model able to simulate the superconductor with its real
thickness. Practical CC gaps are in the range between \SI{0.1}{} and
\SI{1}{\milli\meter}~\cite{van_der_Laan:SUST2019,Nguyen:JAP23}, where
no scaling rules can be directly applied to a simulation performed
with artificially expanded thickness.

\section{\label{s:conclusion}Conclusion and Outlook}

We have studied a CORT cable made of one layer of three CCs wound
around a former of radius $R - d/2$ with a pitch angle $\alpha$ and a
separation gap $g$. The current distribution and AC losses were
investigated with the help of an helicoidal finite element model,
exploiting the helical symmetry of the cable, with a power-law model
and a field-independent critical current.

In this helical arrangement of CCs, the current distribution is
non-trivial and was found to  follow the predicted current
Garber pattern over most of the cable cross section, except near the
gap edges where the current density aligns along the CC edge face. As a
result, there is a rich interplay between several loss
mechanisms. The edge losses arise in the vicinity of the gap region
and are predominant for gaps higher than \SI{100}{\micro\meter},
where they approach the prediction from the Mawatari model. The
losses are generated by the perpendicular component of the magnetic
flux density rushing in from the edge towards the centerline of the
CC as current is raised. In contrast, the surface losses are
predominant for the smallest gaps $g \ll \SI{100}{\micro\meter}$ and
are due to the penetration of the tangential components of the
magnetic flux density from the top and bottom surfaces of the
CC. These losses are following the predictions of the surface loss
model, with however a large sensitivity with respect
to the value of $n$. 
 
The Garber current pattern predicts the occurrence of current lines
oriented azimuthally on the inner face of the CCs and axially along
their outer face. These two current flows have to be connected to one
another near the gap ends of the CCs by diving currents. However,
  the radial component of the current density is much smaller than the
  azimuthal and axial components and their contribution to the AC
  losses is thus negligible, given the power law resistivity.  A further point concerns the subcriticality
of $J_r$: it is indicative of low radial electric fields and is
stronger for low values of $n$. This behavior also appears to be more
pronounced for thin CCs, a property that is reminiscent of the
electric field scaling observed in a coaxial tubular conductor, as
discussed in Appendix~\ref{A:subcritical}.

Comparing losses in CCs with different thicknesses shows that working
with artificially expanded CCs and using a scaling law only applies
(with some approximation) for the smallest gaps. In addition, AC
losses for cases of intermediate gaps $g \sim \SI{100}{\micro\meter}$
cannot be estimated with analytical expressions. These results show
the importance of having a numerical model that can simulate CCs with
their real geometrical parameters.  Practical CORT cables have gaps in
the range between 0.1 and \SI{1}{mm}, an interval where both the
surface loss model and the Mawatari model underestimate the
losses. This study also underlines the importance of the Garber
current pattern, which necessitates modeling the CCs over their
thickness. In cables with multiple layers, it was shown in
\cite{Nguyen:JAP23} that the edge losses can be reduced by overlapping
successive layers without aligning the gaps, while the losses can also
be reduced for individual layers by overlapping adjacent
CCs~\cite{Siahrang:SUST2011}. In such circumstances, it is expected
that the Garber losses will be predominant. A model capable of
reproducing the actual current pattern (which in multilayer cables
will be different from that treated in this work,
see~\cite{Clem:SST10}) and the losses is therefore needed.
 
This study left out a number of effects: the current was chosen in
such a way that the magnetic flux fronts penetrating from the inner
and outer faces do not cross. The flux-cutting regime that would
result from an overlap of the flux fronts will be studied in a future
work. This study also did not address the dependence of the critical
current density with the magnitude and direction of the magnetic flux
density. The Garber pattern is expected to still be present with
modified $J_{\rm c}(B)$ laws, since it is arising from the helical
symmetry of the cable. However, the effects of the value of $n$
might be softened~\cite{Vanderbemden:PRB07}. Last, the anisotropy of
the critical current density in the CC was neglected. This property
has the potential to increase the relative contribution of the diving
current losses.  

\appendix
\subsection{\label{A:helicoidal-model} Helicoidal model}

\begin{figure}
  \centering
  \includegraphics[width=0.65\linewidth]{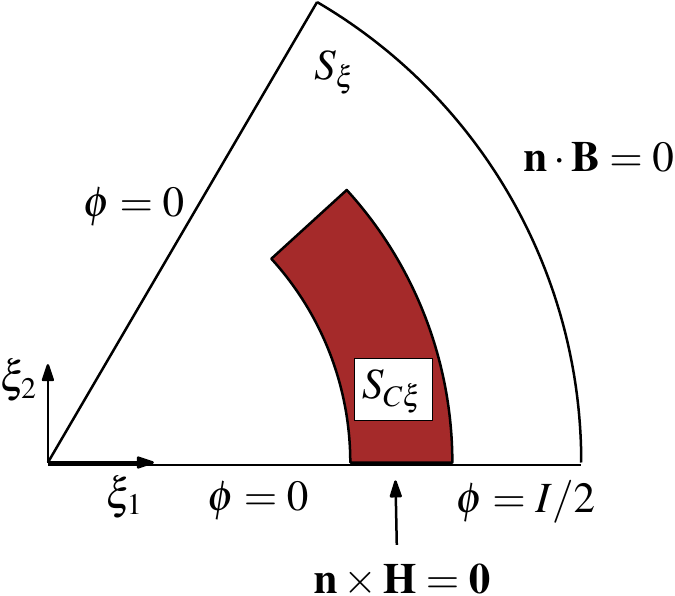}
     \caption{Geometry and boundary conditions of the helicoidal model
       in the helicoidal coordinate system. With equal currents
       flowing in each CC, only $1/6^{\rm th}$ of the geometry is
       modelled. $S_{C\xi}$ is the superconducting domain and $S_\xi$
       is the complementary domain. The scalar magnetic potential
       $\phi$ has a support in $S_\xi$ and on the boundary of
       $S_{C\xi}$. Current is imposed via essential conditions on
       $\phi$ as shown, while a symmetry condition is imposed in the
       conductor along the $\xi_1$ axis. A natural boundary
       condition is imposed at
       a far radial distance for the in-plane field.}
      \label{f:helicoidal-model-geometry}
   \end{figure}

The helicoidal model is solved in a coordinate system such that the
components of the 3D magnetic field can be discretized with basis
functions allowing the description of any helically symmetric magnetic
fields and their associated current distribution~\cite{Dular:TAS24}.
As each CC carries an identical current, the model can be further
simplified to a sixth of the full cross section as shown in
Fig.~\ref{f:helicoidal-model-geometry}, in which the boundary
conditions are also specified.

   For investigating the current distributions, a fine mesh was used
   for the CC (due to the sixfold symmetry, only half a CC was
   modelled). It was divided into two regions: The first region
   extends from the gap edge at $\theta = \beta$ to lower values of
   $\theta$ and has a structured trapezoidal mesh made of
   $20 \times 20$ elements of thickness $\SI{0.05}{\micro\meter}$ and
   length $\SI{0.25}{\micro\meter}$. The second region contains
   $20 \times 400$ elements of thickness $\SI{0.05}{\micro\meter}$ and
   lengths that increase as $\theta \to 0$ following a geometric
   progression.  A coarser mesh is used for the AC loss
   calculations. The CC is composed of a region of $20\times 10$
   elements of thickness $\SI{0.05}{\micro\meter}$ and length
   $\SI{0.5}{\micro\meter}$. The second region contains
   $20 \times 200$ elements of thickness $\SI{0.05}{\micro\meter}$ and
   lengths that increase as $\theta \to 0$ following a geometric
   progression. For the thickness scaling study, the same parameters
   are used for $d=\SI{4}{\micro\meter}$ and
   $d=\SI{10}{\micro\meter}$, while for $d=\SI{40}{\micro\meter}$, a
   mesh with 160 elements along the azimuth is used. In each case,
   there are 20 elements across the thickness. In all studies, the
   non-conducting domain has an unstructured mesh with the size of
   elements controlled by a Gmsh size field (with a threshold
   guaranteeing fine mesh elements in the air near the gap edge up to a
   distance of a few microns) and the circular outer boundary is
   situated $\SI{3}{\cm}$ away from the center of the cable.

\subsection{\label{A:subcritical} Subcritical current density in a thin tube transporting a current $I < I_{\rm c}$ }

In the course of the study, it was observed that the current density
components were frequently observed to be subcritical, to a level that
is more pronounced as the thickness $d$ decreases. Together with this
behavior, the current distribution was found to be sensitive to 
the $n$-value. These observations are reminiscent of a result in the
textbook of Carr~\cite{Carr}, stating that for a long wire carrying a
current $I$ varying sinusoidally over time, the axial electric field
induced within the layer being penetrated is given by
\begin{equation}
 \label{eq:Carr}
   E_z(r) = \frac{\mu_0}{2 \pi} \,\frac{d I}{d t}\,\log\frac{r}{R_1},
\end{equation}
where $R_1$ is the position of penetration front and
$R_1 < r < R_{\rm out}$ with $R_{\rm out}$ the radius of the
wire. This result can be transposed to the case of a tube of external
radius $R_{\rm out}$ and internal radius $R_{\rm in}$ transporting an
axial current $I < I_{\rm c}$ and thus being partially penetrated with
$R_{\rm in} \leq R_1$. Defining $R$ as the mean radius
\begin{equation}
  \label{eq:mean-radius-Carr}
  R = \frac{R_{\rm out}+R_{\rm in}}{2},
\end{equation}
one can express $r$ and $R_1$ as
\begin{equation}
  \label{eq:R_1-r}
   r = R + a\,\frac{d}{2} \quad R_1 = R + b\,\frac{d}{2}, 
\end{equation}
where $d$ is the tube thickness, $a \in [0,1]$ is a numerical constant 
depending on $r$ and $b \in [0,1]$ is a constant depending on the net current
$I$.  Performing a series expansion of the logarithm for small $d/R$ in
\eqref{eq:Carr} then gives $E_z(r) \propto d$ for any
$r$ in the penetration front.

This result is established in the limit of the critical state model,
but can also be observed numerically at finite $n$. As a result, from
the power law, one finds that $J(r)/J_{\rm c} \propto (d/R)^{1/n}$. Hence,
as the thickness decreases, the current $J$ becomes subcritical and
the effect is more pronounced for smaller values of $n$.

\section*{Acknowledgments}
The authors thank Raphaelle Chan (KIT) for proofreading the manuscript
and for helpful comments on the text. Computational resources were
provided by the Consortium des Équipements de Calcul Intensif (CÉCI),
funded by the Fonds de la Recherche Scientifique de Belgique
(F.R.S.-FNRS) under Grant No. 2.5020.11 and by the Walloon Region.

\bibliographystyle{IEEEtran}

\bibliography{Garber_paper_biblio}

\end{document}